\documentclass[a4paper,10pt,twoside]{cpc-hepnp}

\usepackage{multicol}
\usepackage{graphicx}
\usepackage{graphics}
\usepackage{booktabs}
\usepackage{amssymb,bm,mathrsfs,bbm,amscd}
\usepackage[tbtags]{amsmath}
\usepackage{lastpage}

\begin{document}


\fancyhead[c]{\small Submitted to Chinese Physics C }
\fancyfoot[C]{\small 010201-\thepage}


\footnotetext[0]{Received  March 2013}

\title{Study on optimization of water Cherenkov detector array of LHAASO project for surveying VHE gamma rays sources}

\author{%
\quad LI Hui-Cai $^{1;2}$\email{lihuicai@ihep.ac.cn}%
\quad CHEN Ming-Jun $^{2}$
\quad JIA Huan-Yu $^{1}$
\quad GAO Bo $^{2}$
\\\quad WU Han-Rong $^{2}$
\quad YAO Zhi-Guo $^{2}$
\quad YUO Xiao-Hao $^{2;3}$
\quad ZHOU Bin $^{2}$
\quad ZHU Feng-Rong $^{1}$
}

\maketitle

\address{%
$^1$ School of Physical Science and Technology, Southwest Jiaotong University, Chengdu, 610031, China\\
$^2$ Institute of High Energy Physics, Chinese Academy of Sciences, Beijing 100049, China\\
$^3$ Normal University of Hebei, Shijiazhuang, 050016, China\\
}

\begin{abstract}
A water Cherenkov detector array, LHAASO-WCDA, is proposed to be built at Shangri-la, Yunnan Province, China. As one of the major components of the LHAASO  project, the main purpose of it is to survey the northern sky for gamma ray sources in the energy range of 100~GeV¨C 30~TeV. In order to design the water Cherenkov array efficiently to economize the budget, a Monte Carlo simulation is proceeded. With the help of the simulation, cost performance of different configurations of the array is obtained and compared with each other, serving as a guide for the more detailed design of the experiment in the next step.
\end{abstract}




\begin{multicols}{2}

\section{Introduction}

The detection of Very-High-Energy (VHE, $>100$ GeV) cosmic gamma rays has been campaigned vigorously since the first detection of TeV gamma radiations from the Crab Nebula by the Whipple experiment in 1989~\cite{lab1}. Two major detection approaches exist in this research field: Imaging Air Cherenkov Telescopes (IACTs)~\cite{lab2} and ground particle detector arrays~\cite{lab3}. The formers win in angular resolution ($<0.1^\circ$) and background rejection power so that they possess a better sensitivity in morphology observation and spectrum measurement. However, because of low duty-circle ($\sim$10\%) and narrow field of view ($<5^\circ$ typically), their capability in full sky survey and long-term monitoring of sources is very limited. This aspect is fortunately complemented by ground particle arrays, which can work all the time (duty cycle $>95\%$) and monitor a large piece of the sky ($>$2$\pi$/3) simultaneously, as shown by precedent experiments such as Tibet AS$\gamma$, Milagro, and ARGO-YBJ. One of the techniques used in this kind of approach, the water Cherenkov, has the unique advantage of much better background rejection power than other options like plastic scintillator and RPC, which is well demonstrated by simulations and has verified by Milagro experiment.

Targeting gamma astronomy in energy band from 100~GeV to 30~TeV, the water Cherenkov detector array (WCDA)~\cite{lab4} of the Large High Altitude Air Shower Observatory (LHAASO)~\cite{lab5}, covering an area $90,000\rm\;m^2$, has been proposed to be built at Shangri-la( 4300~m a.s.l.), Yunnan, China. Much upon the experience of Milagro experiment, the current official configuration of WCDA is 5~m $\times$ 5~m cell-sized, and incorporated a single photomultiplier at a water depth of 4~m for each cell~\cite{lab4}.  The simulation of the array in this configuration shows a very good performance in sensitivity to gamma ray sources, particularly at the energy band around 5~TeV. But it is not necessarily the best configuration, for instance for gamma rays at low energies, and especially when the cost factor is taken into account. This paper is just to carry out this study, to see what the best configuration of the array is in the sense of cost performance, via tuning the cell size, the water depth and the number of PMTs.

In section 2, the scheme of the optimization is introduced, and next the optimization procedures and results based on Monte Carlo simulations are presented, seen in section 3.

\section{Optimization scheme}
To simplify the optimization procedure, some discontinuous values of the detector configuration parameters regarding the cell size, the water depth and the number of PMTs in each cell are proposed to be taken as the starting point of the simulation. Figure 1 shows the sketch drawing of 4 neighboring cell detectors in two view angles, where parameters $L$, $H$, and $N$ are subjected to be tuned. Three cell sizes such as 4~m $\times$ 4~m, 5~m $\times$ 5~m and 6~m $\times$ 6~m are selected; three effective water depths such as 3~m, 4~m and 5~m are chosen; five groups of PMT quantities in each cell such as 1, 2, 3, 4 centered and 4 scattered are adopted - as shown in figure 2, where the verbal adjectives of the last two groups with 4 PMTs represent where the four PMTs are located, near the center or much separated in the cell. Each iteration of these 3 kinds of elements can be combined to form a particular configuration. Altogether there are $3 \times 3 \times 5 = 45$ configurations to be proceeded, whose cost performance are due evaluated and compared.

\begin{center}
\includegraphics[width=0.60\linewidth]{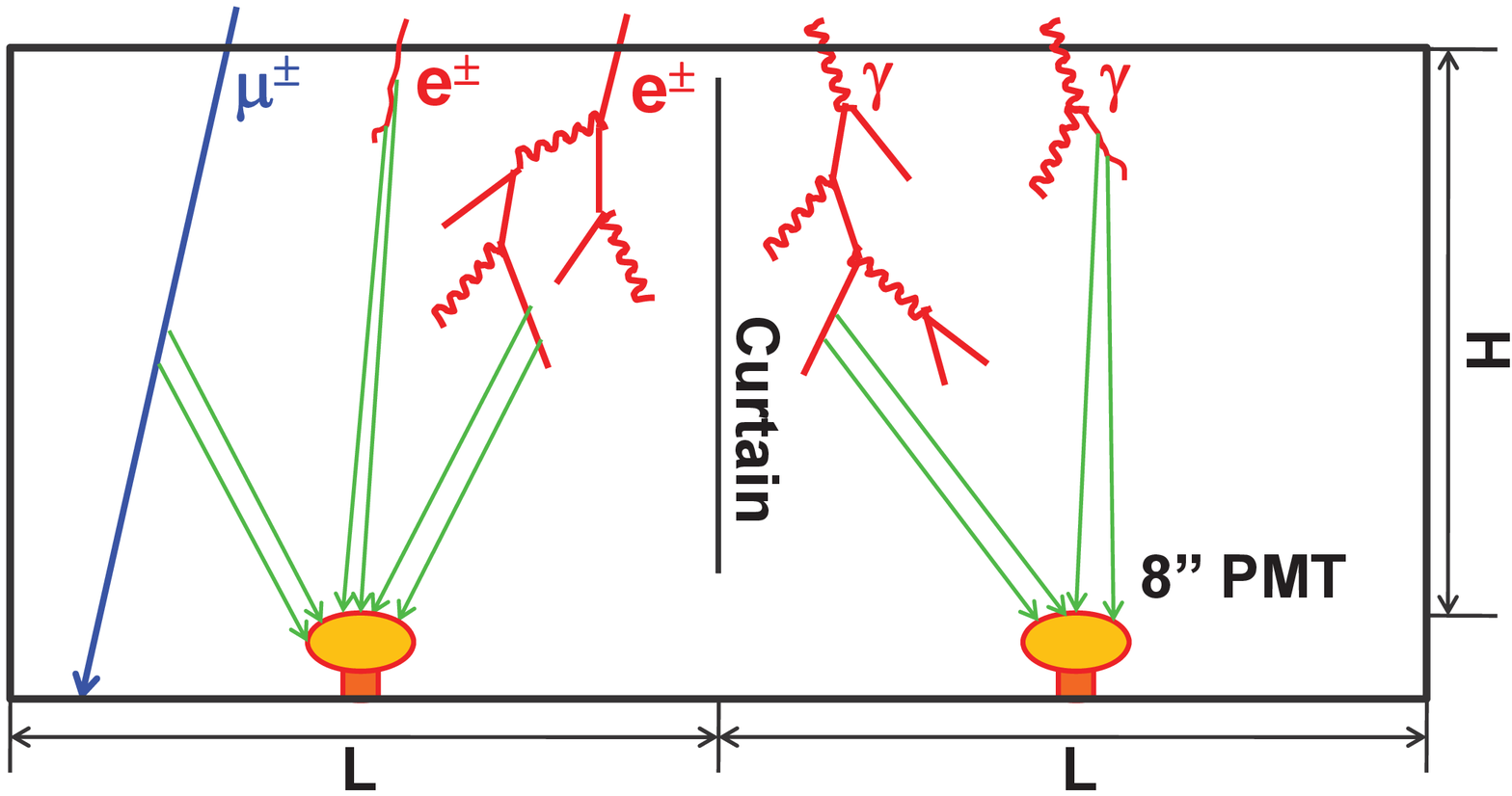}
\includegraphics[width=0.30\linewidth]{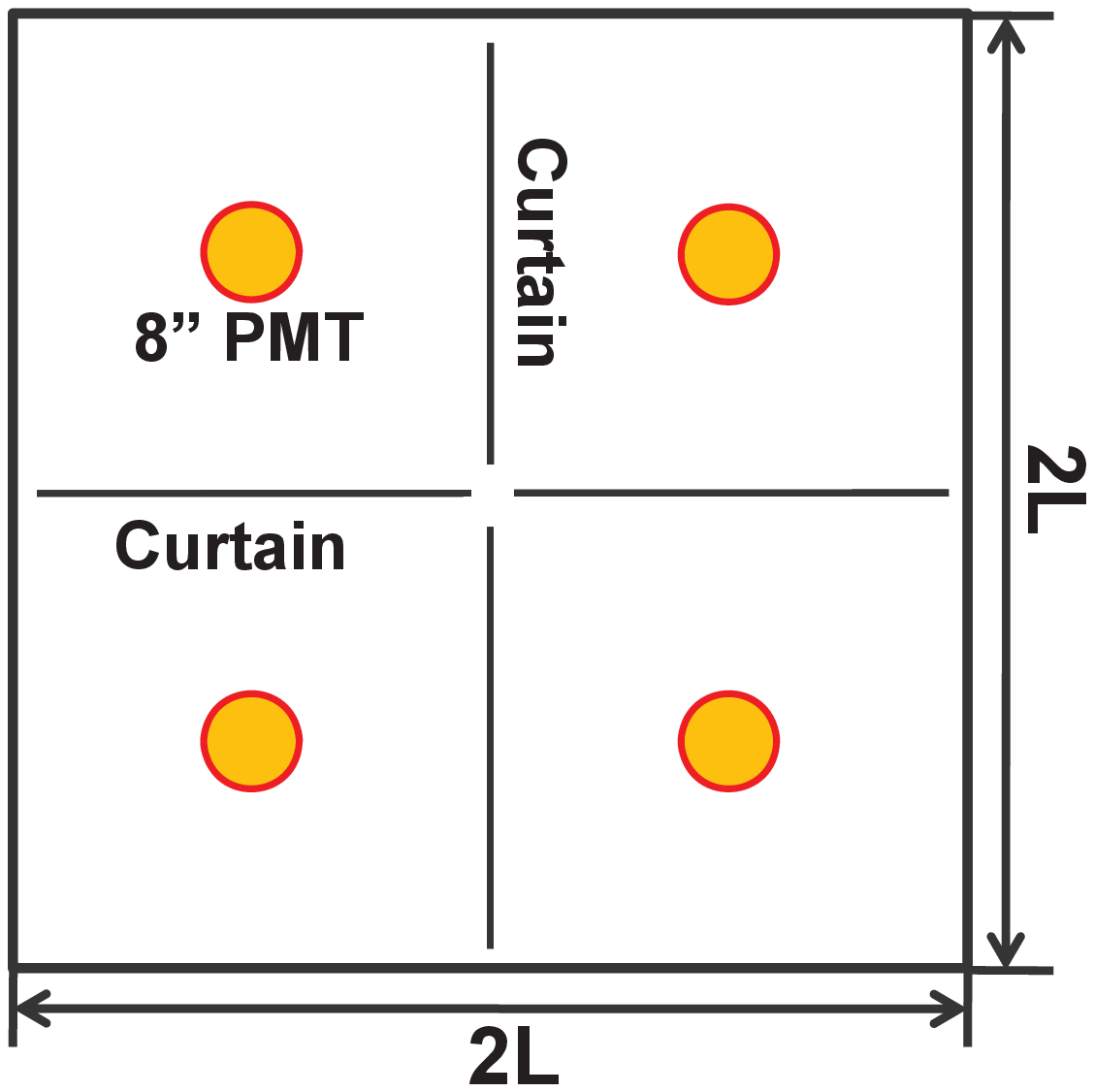}
\figcaption{\label{fig1}Sketch drawing of 4 neighboring cell detectors. Left: side view; Right: top view. Where L is the size of cell and H is the water depth. As a diagramatic plot, only 1 PMT in each cell is drawn, which actually represents $N$ PMTs.}
\end{center}

\begin{center}
\includegraphics[width=0.95\linewidth]{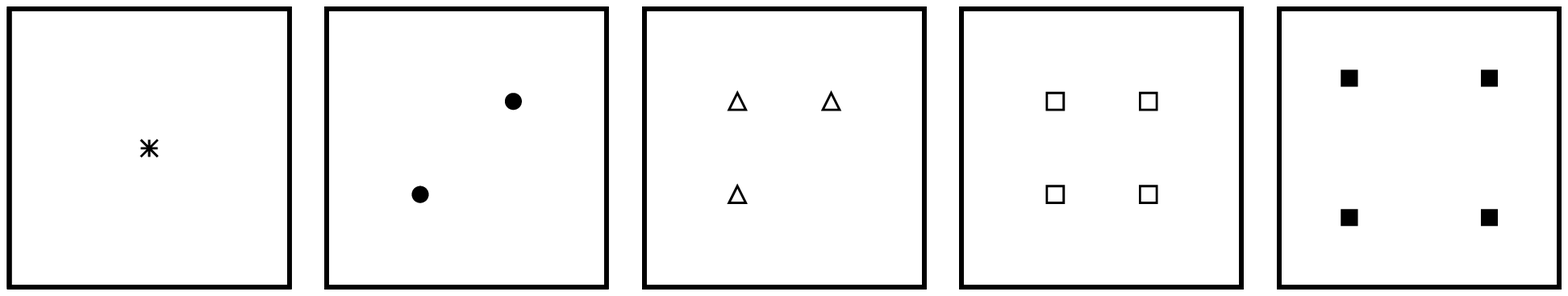}
\figcaption{\label{fig2} 5 groups of PMT quantities in a cell. $\divideontimes$: 1~PMT, $\bullet$: 2~PMTs, $\vartriangle$: 3~PMTs, $\square$: 4~PMTs centered, $\blacksquare$: 4~PMTs scattered.}
\end{center}

\subsection{Simplified simulations}

A full simulation should be the best approach for calculating the performance of each configuration. But in practice it would be exhausted due to the fact that the simulation procedure is quite time-consuming, especially the process of production and tracing of the Cherenkov lights in the water. As to the array of current design, which is just one example of above 45 configurations, 3 months were taken with a PC farm of 100 CPU cores. In order to accomplish the optimization for all above configurations, years of computing time are expected. It is of course not a practical solution. To get over this obstacle, a simple and efficient optimization procedure is adopted and proceeded, as follows.

First of all, in order to lessen the burden of iterations, in the simulation, for each cell, overall 9 PMTs are put into the detector configuration at the same time, see figure 3. These 9 PMTs can be easily classified afterwards in the off line into the 5 groups required for the optimization. This kind of overall configuration would not change the simulation results much, as the area of 9 PMTs is very limited comparing the whole cell bottom, and as they react more-or-less like the surrounding curtains or plastics in black of each cell for the purpose of avoiding reflections of Cherenkov lights. With help of this treatment, the total number of iterations can be reduced to be 3$\times$3 = 9.

\begin{center}
\includegraphics[width=0.80\linewidth]{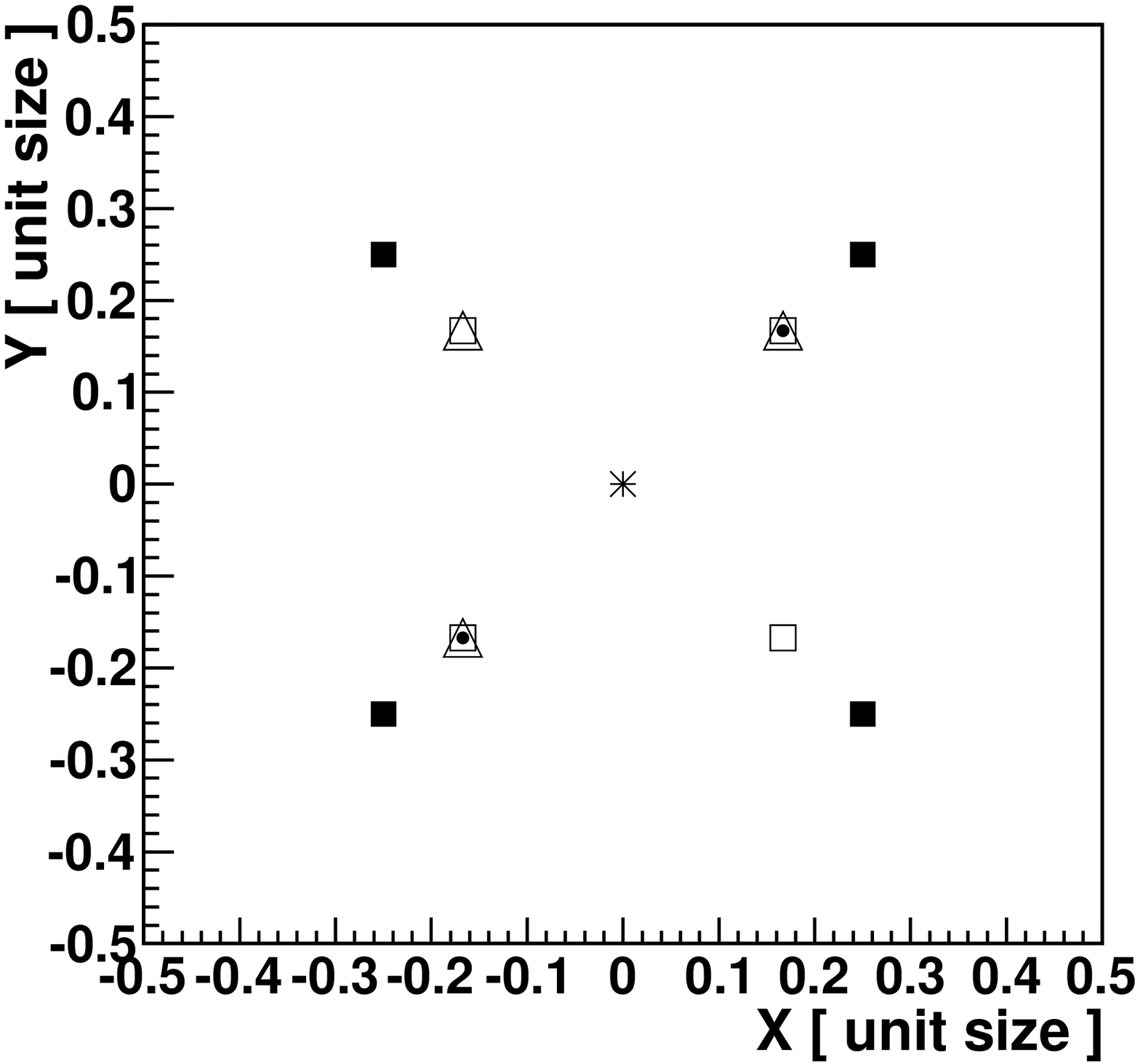}
\figcaption{\label{fig3} PMT locations for simultaneous simulation of 5 groups of PMT quantities.}
\end{center}

Special scenarios, which mean simplified detector setups and  or simplified sets of primary shower parameters like energy, direction and projection area, are adopted in the study for sake of speeding up the simulation. What we concern is the performance comparison between different detector configurations which can be told by simulations in some special scenarios; the absolute performance values for real cases are important too but they are out of the scope of this study. After that, two simulation approaches dealing with two different special scenarios are carried out.

\begin{itemize}
  \item [$i.$] Single cell approach\\
  As the simplest scenario, just a single cell detector is used for the simulation. Secondary particles of hundreds of primary gamma showers are thrown one by one onto the cell detector, and then their interaction and transport processes in the water and in the PMT are simulated. Counting the number of particles that fired at least 1~PMT, compared with the number that thrown onto the cell, the efficiency $\eta$ of the cell detector is obtained. The efficiency for water Cherenkov is usually less than 30$\%$, but ample photons in the shower secondary particles compensate the loss of efficiency when comparing with other type of arrays such as plastic scintillator or RPCs that detect only charged particles. For the full coverage detector, if major part of the shower particles fall into the array, the efficiency value would not differentiate much from above $\eta$ obtained with the simplest case. The performance parameter in this case is then defined as $\sqrt\eta$, as the sensitivity is usually inverse proportional to the square root of the number of hits, especially for showers with scarcely distributed hits at the low energies. Just because of this reason, this approach is principally for low energy gamma ray sources.

  \item [$ii.$] Array approach\\
  As a more complex scenario, the configuration of an array of 150~m $\times$ 150~m is adopted, but simulated with simplified primary particle sources ¨C gammas and protons with fixed energies, vertically incident, and bombarding at the center of the array. Analyses on angular resolution and proton-gamma discrimination are performed, so that somehow realistic performance results for each configuration are obtained. In the circumstance we are interested, usually gamma showers generate 2-3 times more particles at the ground than that of proton with the same energy, so in the proton-gamma discrimination, the proton energy is deliberately chosen to double the energy of its gamma partner, for instance 0.5~TeV gamma versus 1~TeV proton, and 1~TeV gamma versus 2~TeV proton, and so on, see figure 4 for the details.

\begin{center}
\includegraphics[width=0.90\linewidth]{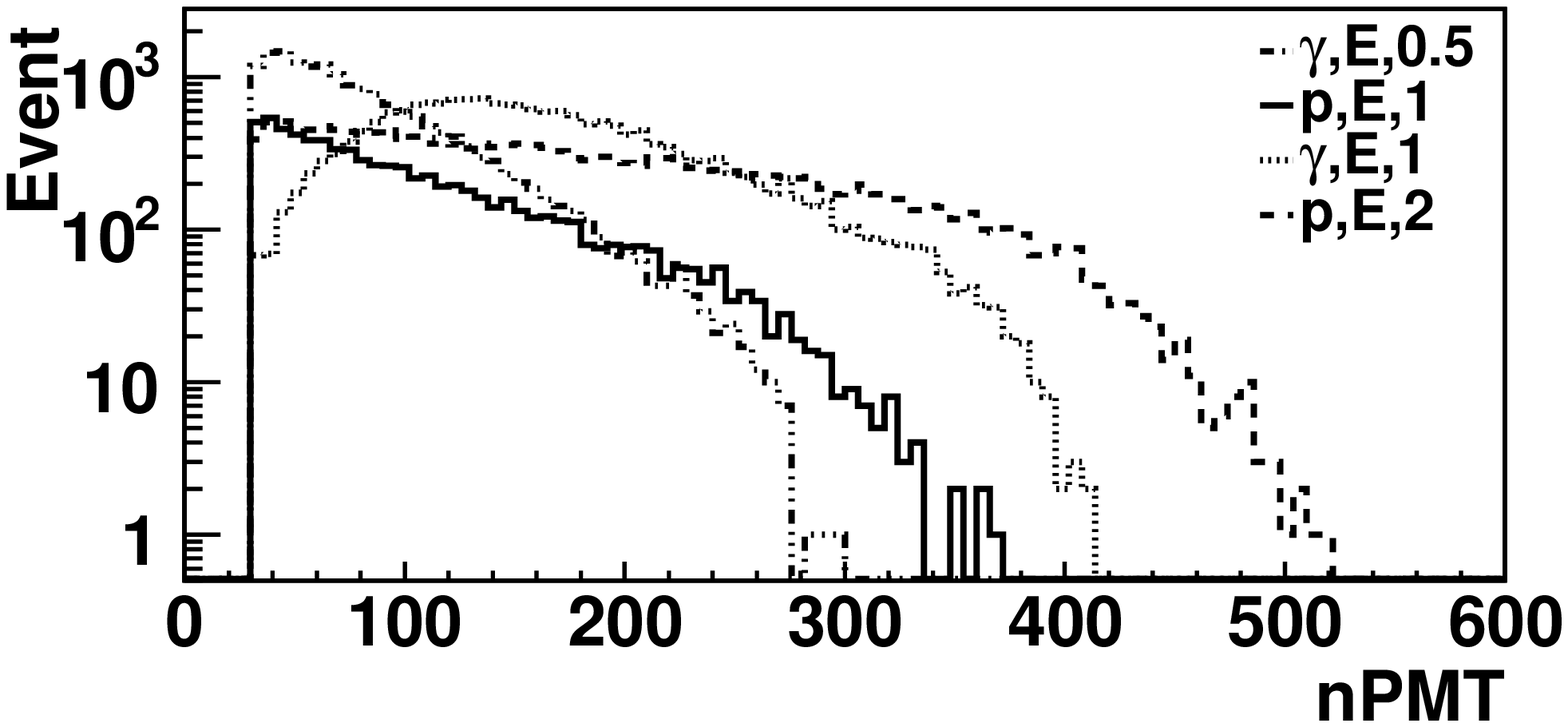}
\figcaption{\label{fig4} Distribution of number of hits in the case of current official detector configuration, for primary gammas and protons at different energies, respectively.}
\end{center}

\end{itemize}

In this study, air shower cascade is simulated with Corsika v6.990~\cite{lab6}, and QGSJETII model~\cite{lab7} is used for high energy hadronic interactions. For avoiding losing the shower information, the kinetic energy cut for secondary particles in Corsika is set to much lower values than that of the Cherenkov production threshold in the water, i.e., 50 MeV for hadrons and muons, 0.3 MeV for pions, photons and electrons. The detector is supposed to be at the altitude of 4300 m a.s.l., and the geometrical setup and the particle propagation are simulated with a code developed on the framework of Geant4 v 9.1.p01~\cite{lab8}.

\subsection{Cost performance}

The cost performance rather than the sole performance of the detector is adopted for the comparison of different configurations. Different configurations with a same array dimension may bring different cost. For example, thicker water depth needs higher pool bank in the construction, larger water volume for filling up, and stronger water purification and recirculation ability in the operating, all of which requires more money; more PMTs in a cell means more PMTs to be ordered and tested, and more electronic channels and cables to be manufactured ¨C that cost more money too. Assuming the detector array is fixed in size of 150~m $\times$ 150~m, the cost of every configuration are carefully evaluated, based on the engineering design reports. The performance of the detector $P$, here defined as the inverse of the flux sensitivity, divided by the cost $C$, the cost performance $P_{\rm C}$ is then calculated, that is
\begin{equation}
\label{four} P_{\rm C}=\frac{\alpha \cdot P}{C}=\frac{\alpha \cdot Q/\theta}{C_{\rm base}+C_{\rm depth}+C_{\rm PMT}}.
\end{equation}
Where $C_{\rm base}$ is the cost of fundamental pool construction including the pond basement and the roof, which actually is a constant in this study, $C_{\rm depth}$ is the cost dependent on the water depth, $C_{\rm PMT}$ is the cost dependent on the total number of PMTs, and $\alpha$ is a constant normalization parameter which sets $P_{\rm C} = 1 $ for the case of the pool in the current official configuration, that is 5~m $\times$ 5~m in cell size, 4~m in water depth, and 1~PMT in each cell. Regardless of the number of PMTs, which is taken account already in $C_{\rm PMT}$, the cell size alone doesn¡¯t affect much on the cost so its contribution is trivial and ignored. Moreover, comparing with $C_{\rm base}$ and $C_{\rm PMT}$, the influence of the water depth to the cost $C_{\rm depth}$ is small.
For approach $i$ in the above, the performance is set to
\begin{equation}
\label{two} P=\sqrt\eta.
\end{equation}
Where $\eta$ is the efficiency above-mentioned. For approach $ii$, the performance is set to
\begin{equation}
\label{three} P=Q/\theta.
\end{equation}
Where $Q$ is the quality factor for proton-gamma discrimination, and $\theta$ is the angular resolution to primary gammas. Those 2 factors are all dependent on the PMT threshold applied in the off line analysis, the $P$ here is the maximum value while ranging the PMT threshold.

\section{Simulation results }

\subsection{Single cell approach}

A detector cell is constructed in the framework of GEANT4. 8-inch PMTs of type Hamamatsu R5912 are used, whose geometrical description and boundary effect code is taken from GenericLAND package~\cite{lab9}. The water absorption length to lights, whose dependence on wavelengths is parameterized according to the curve shape of the measurement of pure water, is scaled to 27~m at 400~nm.

Around 10,000 showers of primary gamma at energy 1~TeV are generated in Corsika. The total number of their secondary particles arriving at the ground amounts to $10^6$, most of which are photons. As to different primary energies, the energy distributions of secondary particles are eventually very close, though actually the energy of primary particle is not important at all in this study. The secondary particles are thrown and tracked in the cell detector one by one, with the particle that fires at least 1~PMT being counted, and finally the efficiency for each configuration is obtained, shown in figure 5. From the plot it is seen that the efficiency is dropping while increasing the cell size, raising the water depth, or reducing the number of PMTs. For the first two trends, it can be explained as the opening angle of PMTs to tracks turning smaller, so that the chance of Cherenkov lights produced along the track hitting the PMT becomes less. The energies of air shower secondary particles are rather low, so usually the tracks of their own or of small showers they initiate appear only at the top part of the water and have scrambled directions. The configuration with water depth lower than 3~m is not simulated, but other studies show that there is a turning point in the efficiency curve around 3~m.
\begin{center}
\includegraphics[width=0.70\linewidth]{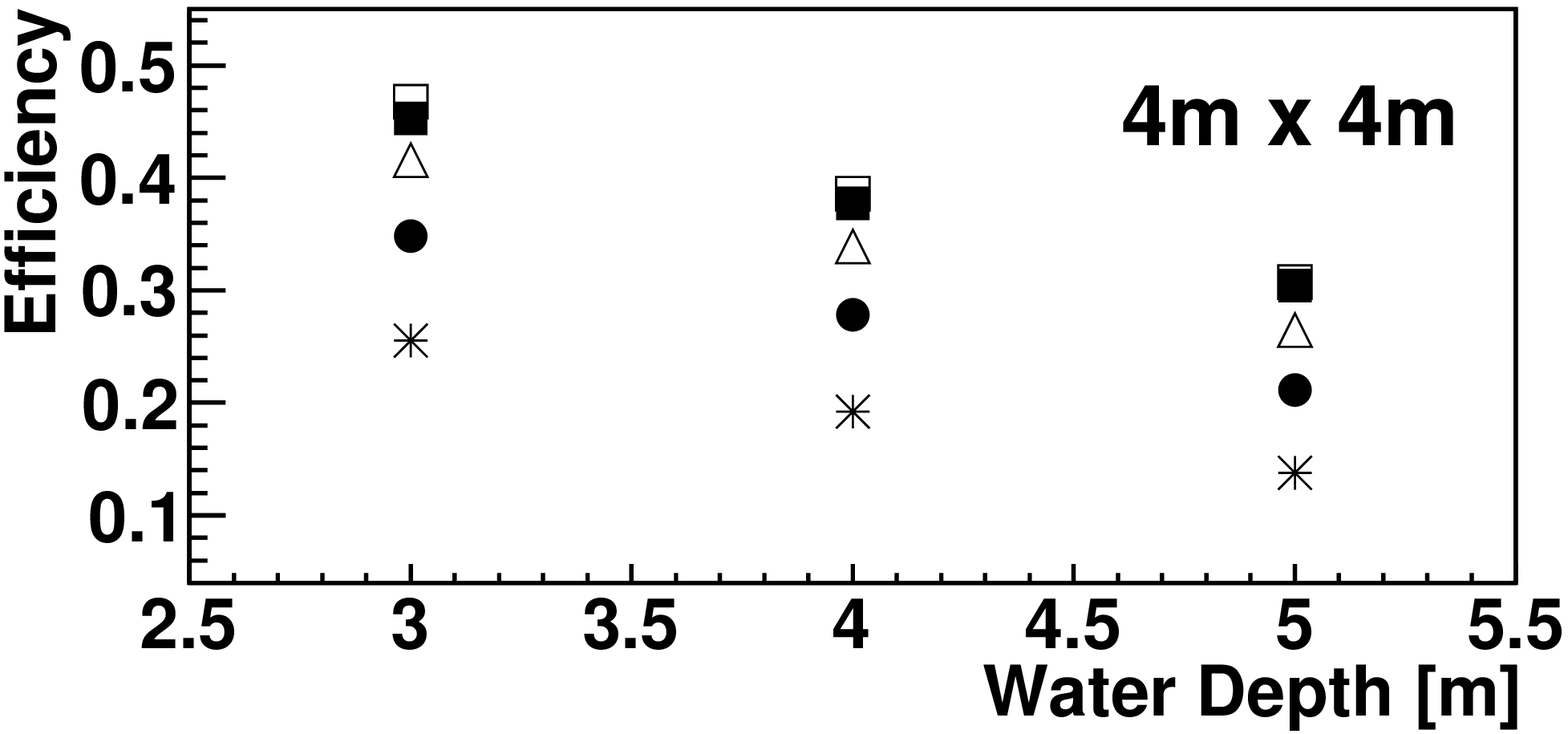}
\includegraphics[width=0.70\linewidth]{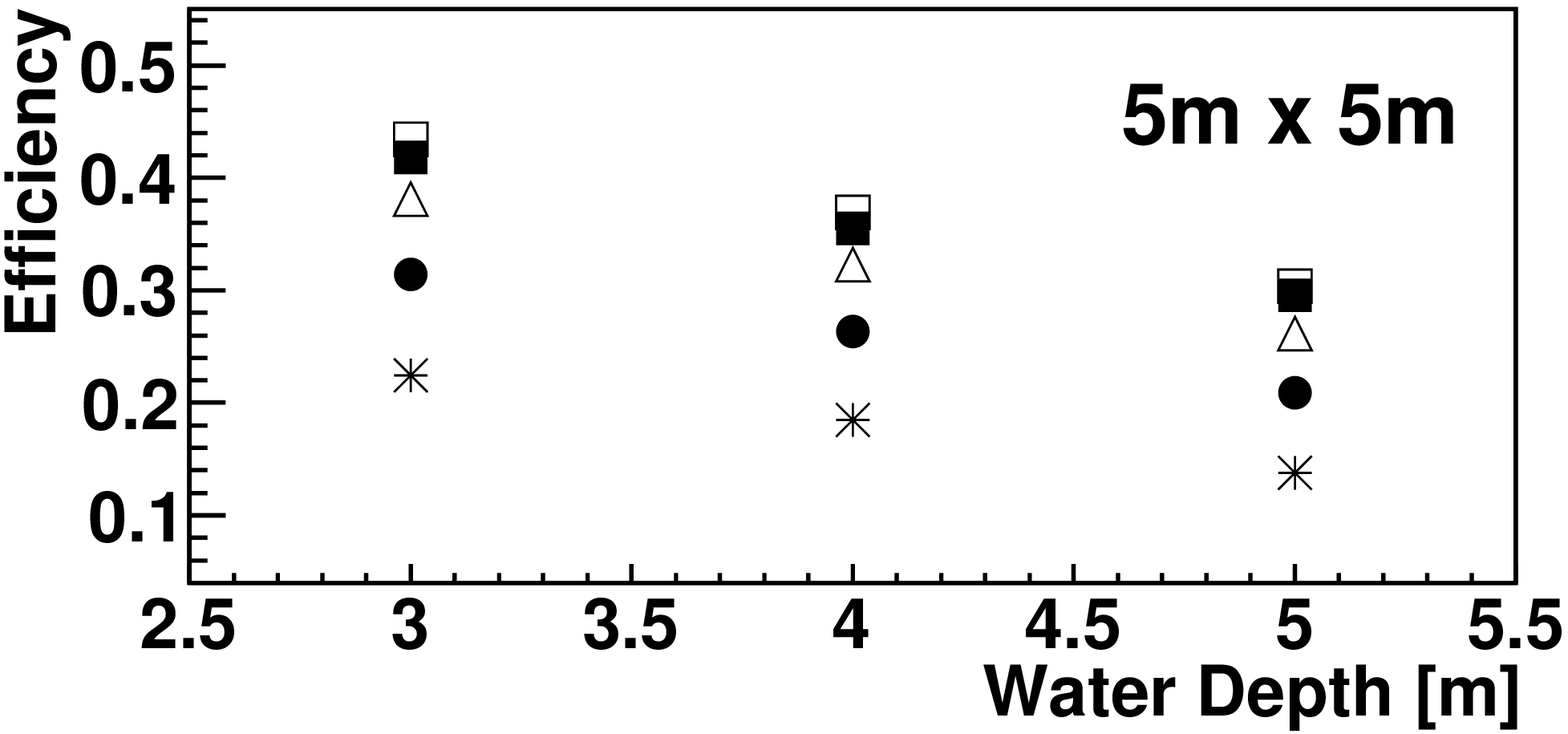}
\includegraphics[width=0.70\linewidth]{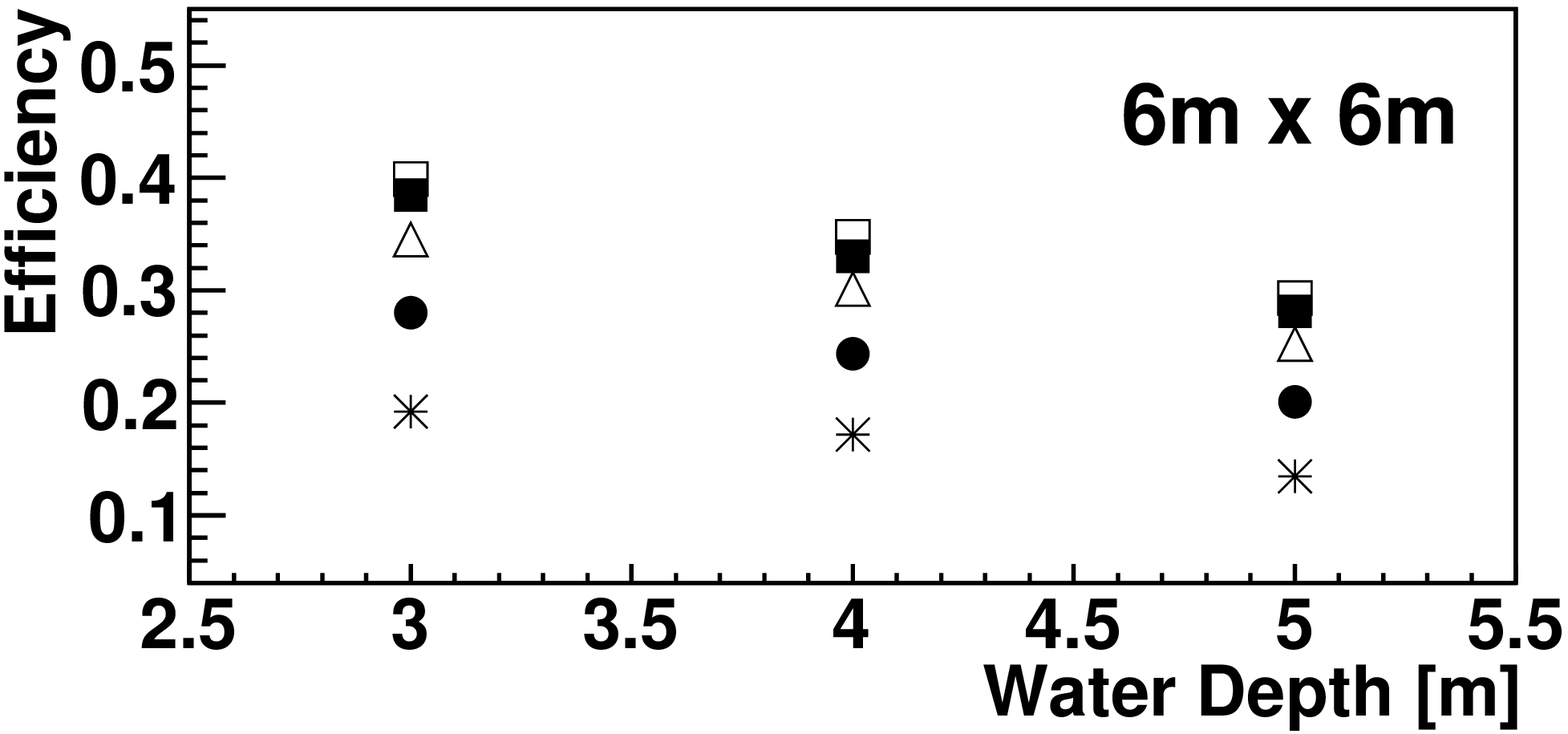}
\figcaption{\label{fig5}  Efficiency of single cell in 45 configurations. Plots in top, middle and bottom are cases for cell sizes of 4~m $\times$ 4~m, 5~m $\times$ 5~m and 6~m $\times$ 6~m, respectively. The abscissa is the water depth, and different marker denote different group of PMT quantities: $\divideontimes$: 1~PMT, $\bullet$: 2~PMTs, $\vartriangle$: 3~PMTs, $\square$: 4~PMTs centered, $\blacksquare$: 4~PMTs scattered. The labels and markers are same for rest figures in this paper otherwise explicitly mentioned.}
\end{center}

The cost of the 150~m $\times$ 150~m array composed of single cells in each of 45 configurations is estimated in the way mentioned in previous section. With equation 1 and 2, the cost performance for each configuration is then obtained, shown in figure 6. It is obvious that the group with single PMT in a cell has better cost performance than that of other groups. As to the water depth, due to the cost changes in the same way as the efficiency, the detector at lower water depth has a better cost performance. It means a shallow water depth is desired for detecting low energy gamma rays. The cell size gains some 20\% too when the cell size is 6~m $\times$ 6~m, because of the reduction of PMT cost.

\begin{center}
\includegraphics[width=0.70\linewidth]{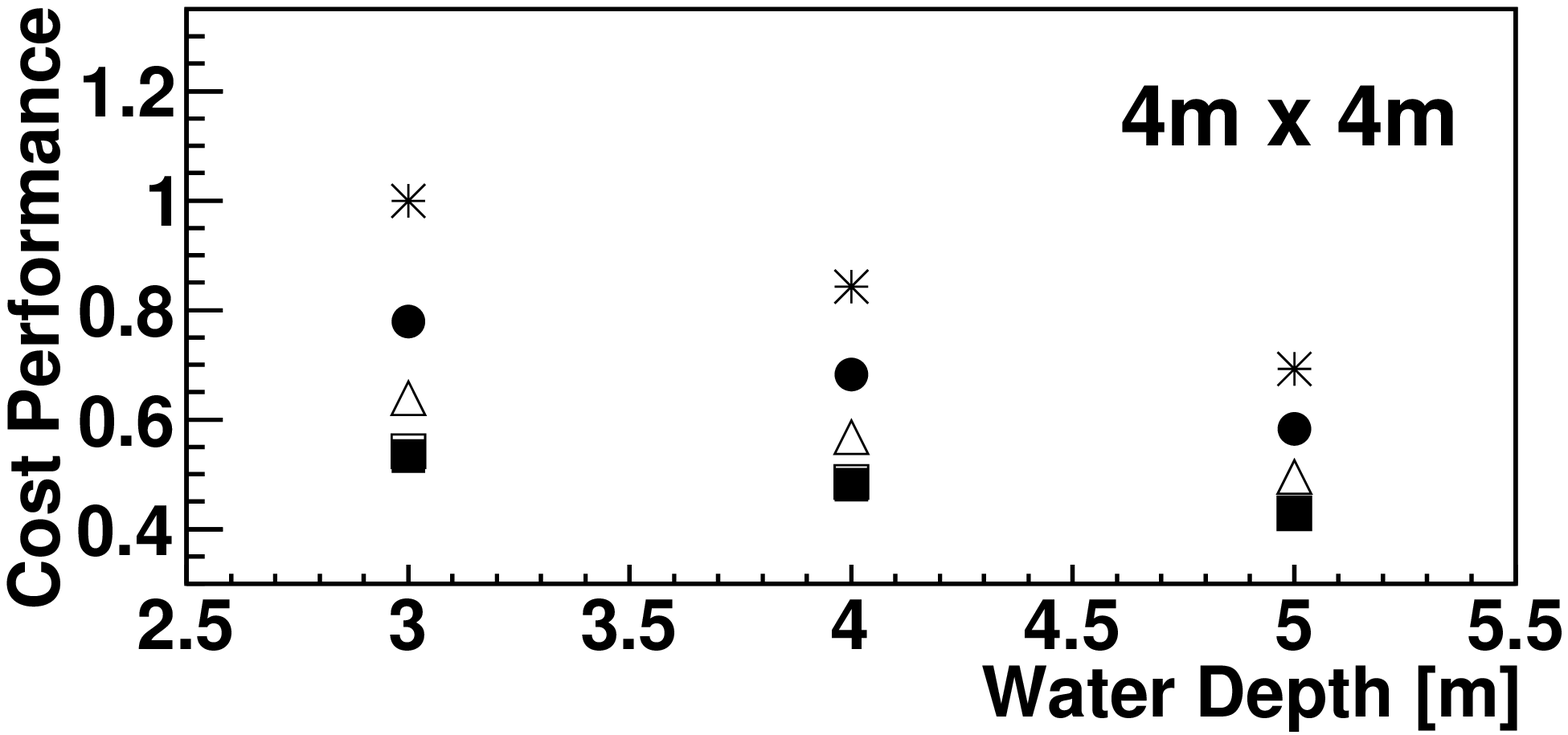}
\includegraphics[width=0.70\linewidth]{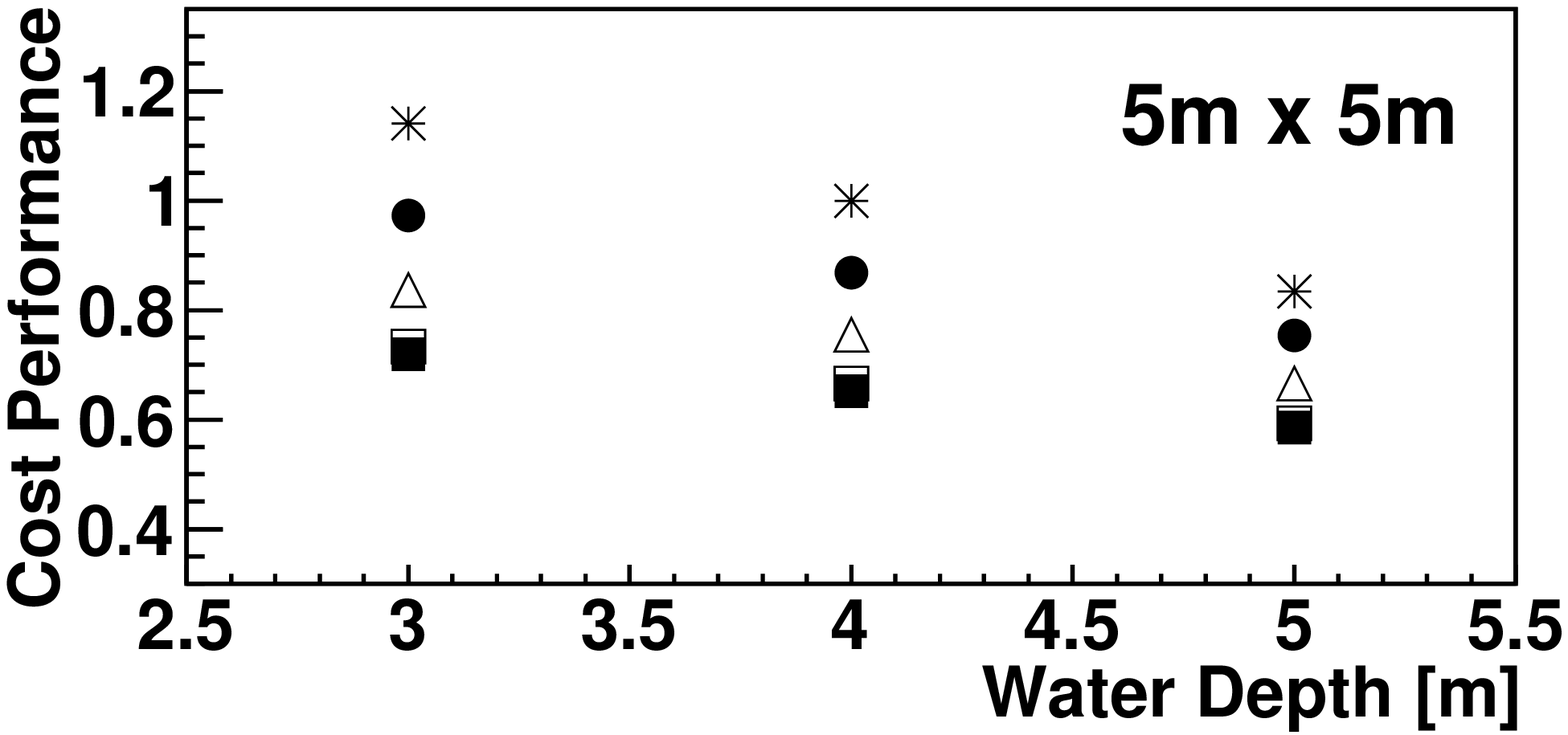}
\includegraphics[width=0.70\linewidth]{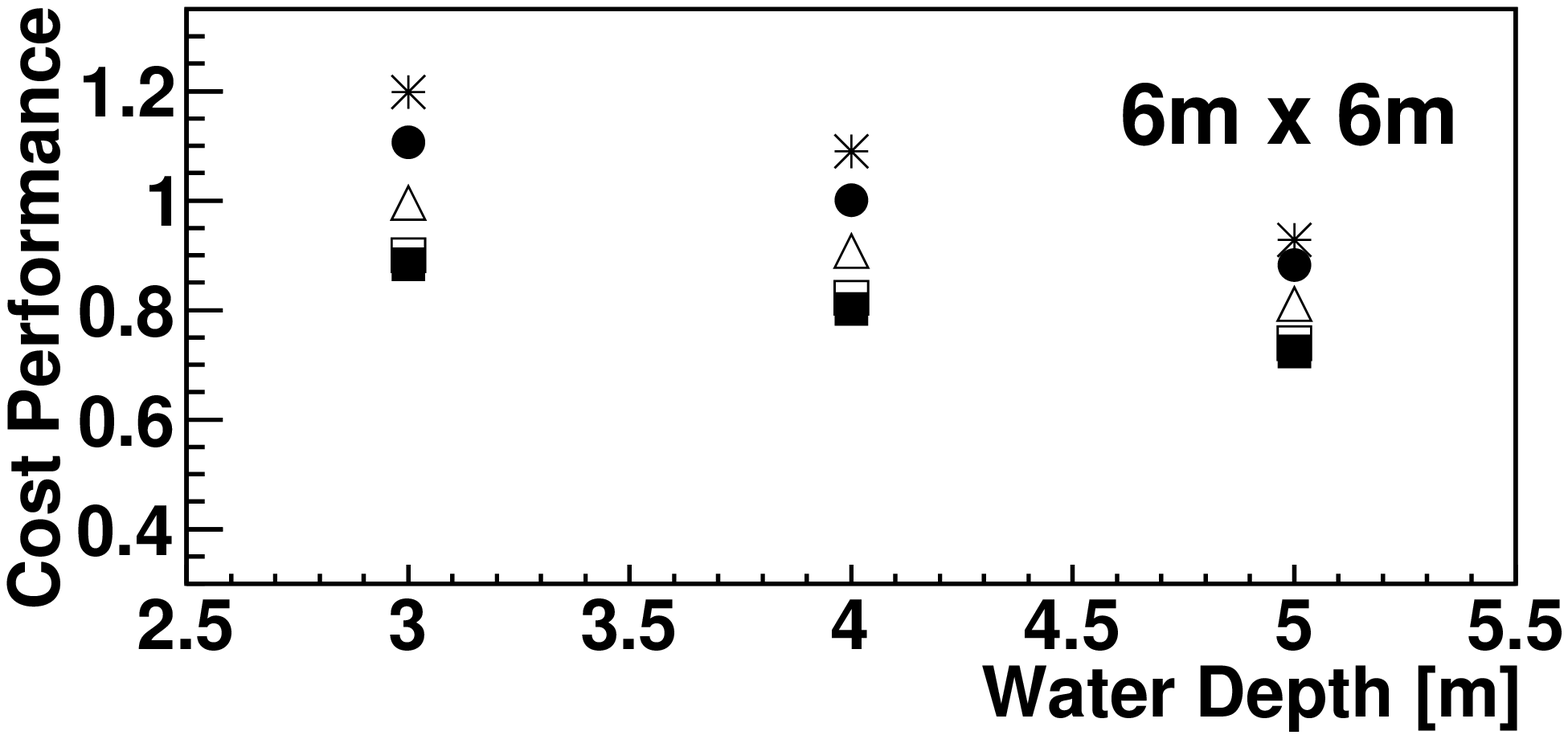}
\figcaption{\label{fig6} Cost performance of the array in 45 configurations with the single cell approach.}
\end{center}

\subsection{Array approach}

In this approach, primary gammas with 3 energies, 0.5, 1 and 2~TeV, are simulated, and that of primary protons, as the background, are set to 1, 2 and 4~TeV correspondingly. For brevity, only the case of 1~TeV gamma and 2~TeV proton is introduced here. As mentioned above, the showers are vertically incident, with core projected to the center of the array. Around 20,000 gamma showers and more proton showers are simulated.

To further lessen the $P_{\rm C}$ express, the configuration with 4~PMTs scattered is abandoned. Both simulations on E/M particles and muons show no obvious differences.

\subsubsection{Gamma proton discrimination}

Water Cherenkov array has an extraordinary property in Gamma proton discrimination. Muons or sub-cores in hadronic showers can produce unevenly lateral distribution of hit intensity on PMTs. When the brightest PMT outside a certain range (e.g., 45~m in radius) of the reconstructed core is chosen, using whose inverse signal amplitude $(1/cxPE)$ to weight the number of fired PMTs $(nPMT)$, the resulting distribution of compactness (= $nPMT/cxPE$) between gamma and proton turns quite different, especially at high energies. Tuning the threshold of compactness, the best quality factor ($Q_{\rm max}$), defined as
\begin{equation}
\label{three}
Q_{\rm max}=\frac{\xi_{\rm s}}{\sqrt{\xi_{\rm b}}},
\end{equation}
is found, where $\xi_{\rm s}$ and $\xi_{\rm b}$ are retained fractions of gamma and proton respectively. Figure 7 shows the $Q_{\rm max}$ for these 36 configurations. The prominent phenomenon is that $Q_{\rm max}$ turns bigger when water depth is higher. The reason behind is as follows: When water depths raises, the cxPE of proton formed most by muon signals suffers little from the water depth hence keeps constant, but $nPMT$ turns smaller due to the dropping of efficiency as shown in previous single cell approach; Gamma shower has very few penetrating muon particles, whose $cxPE$ produced mainly by soft components changes in the way very similar as $nPMT$, so the ratio $nPMT/cxPE$ varies very little.

\begin{center}
\includegraphics[width=0.70\linewidth]{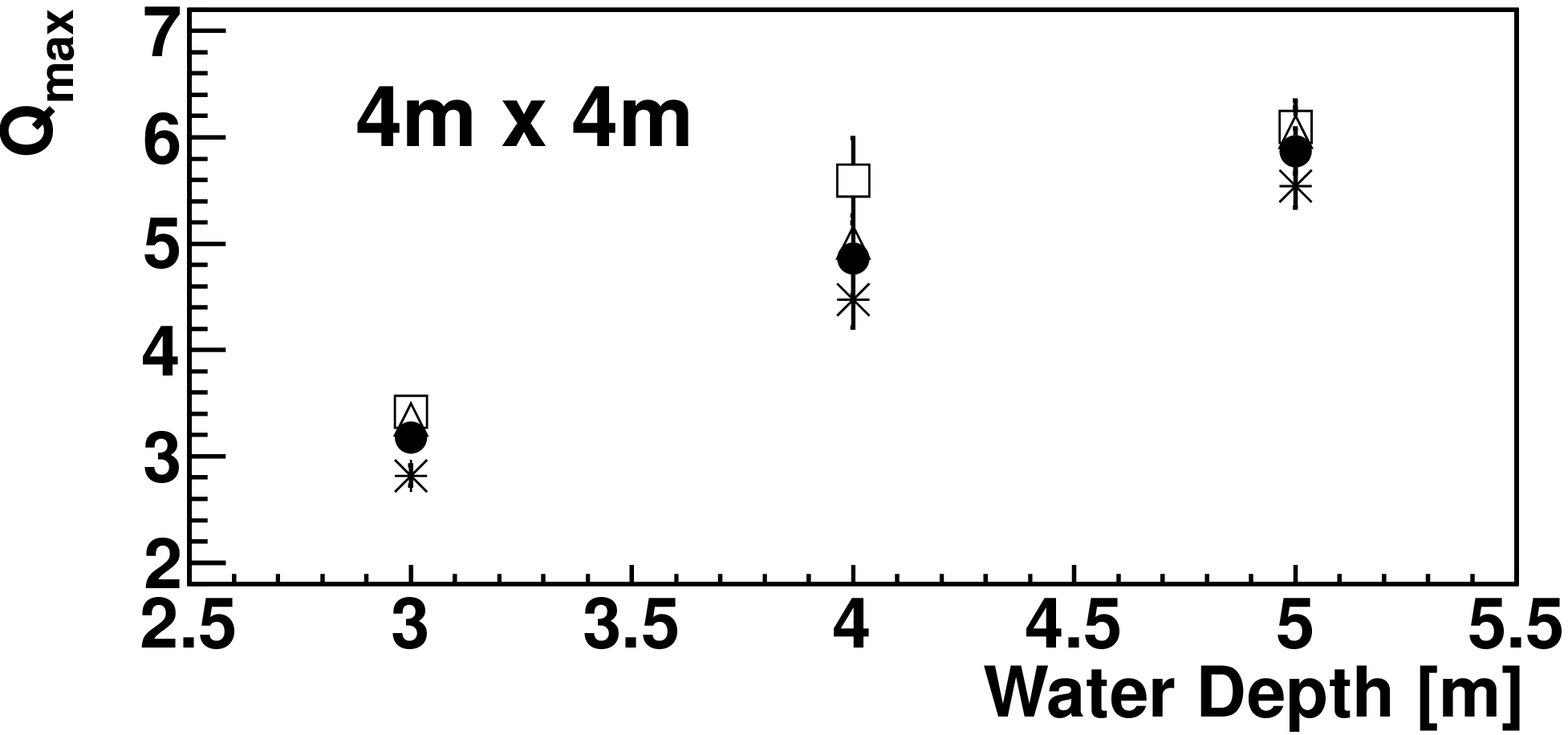}
\includegraphics[width=0.70\linewidth]{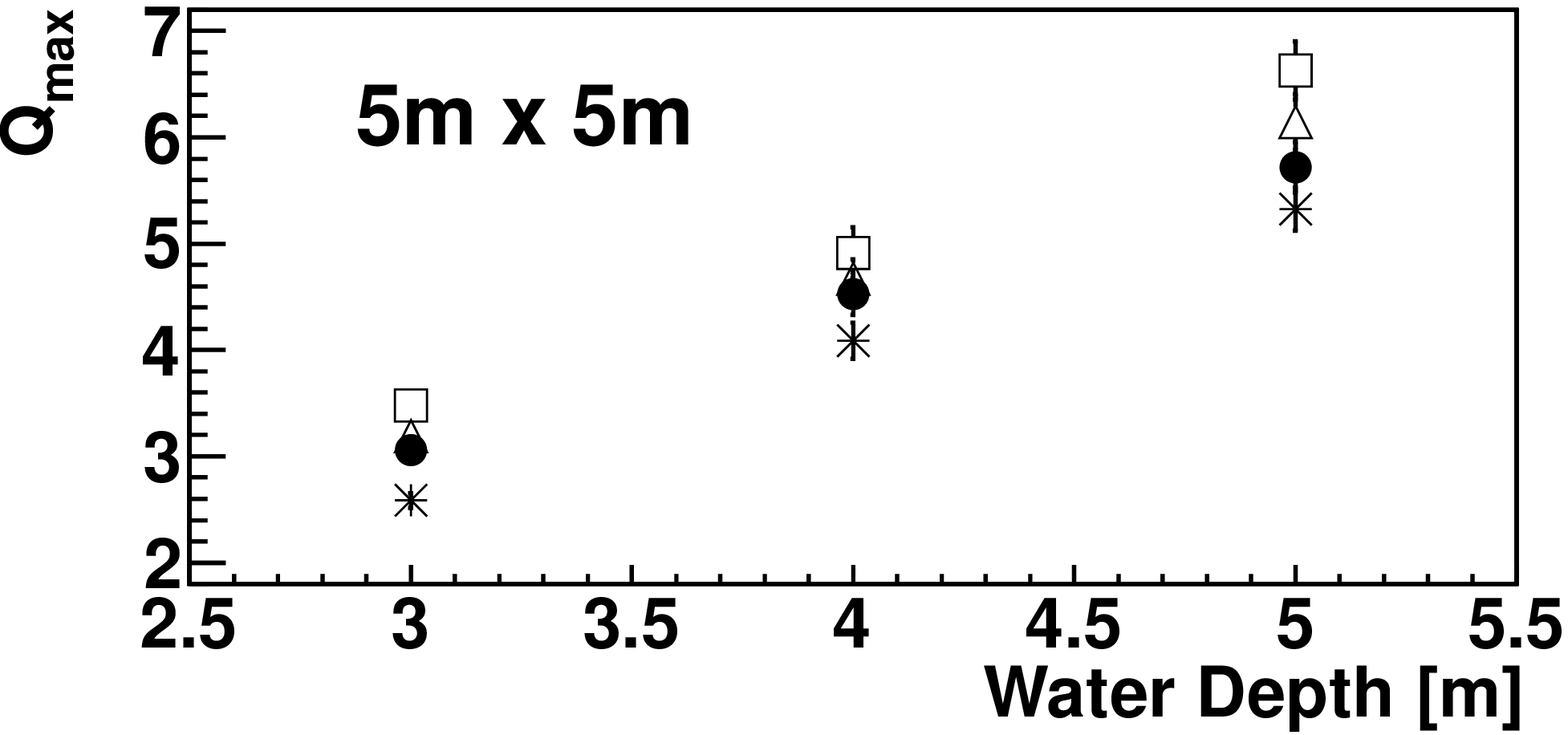}
\includegraphics[width=0.70\linewidth]{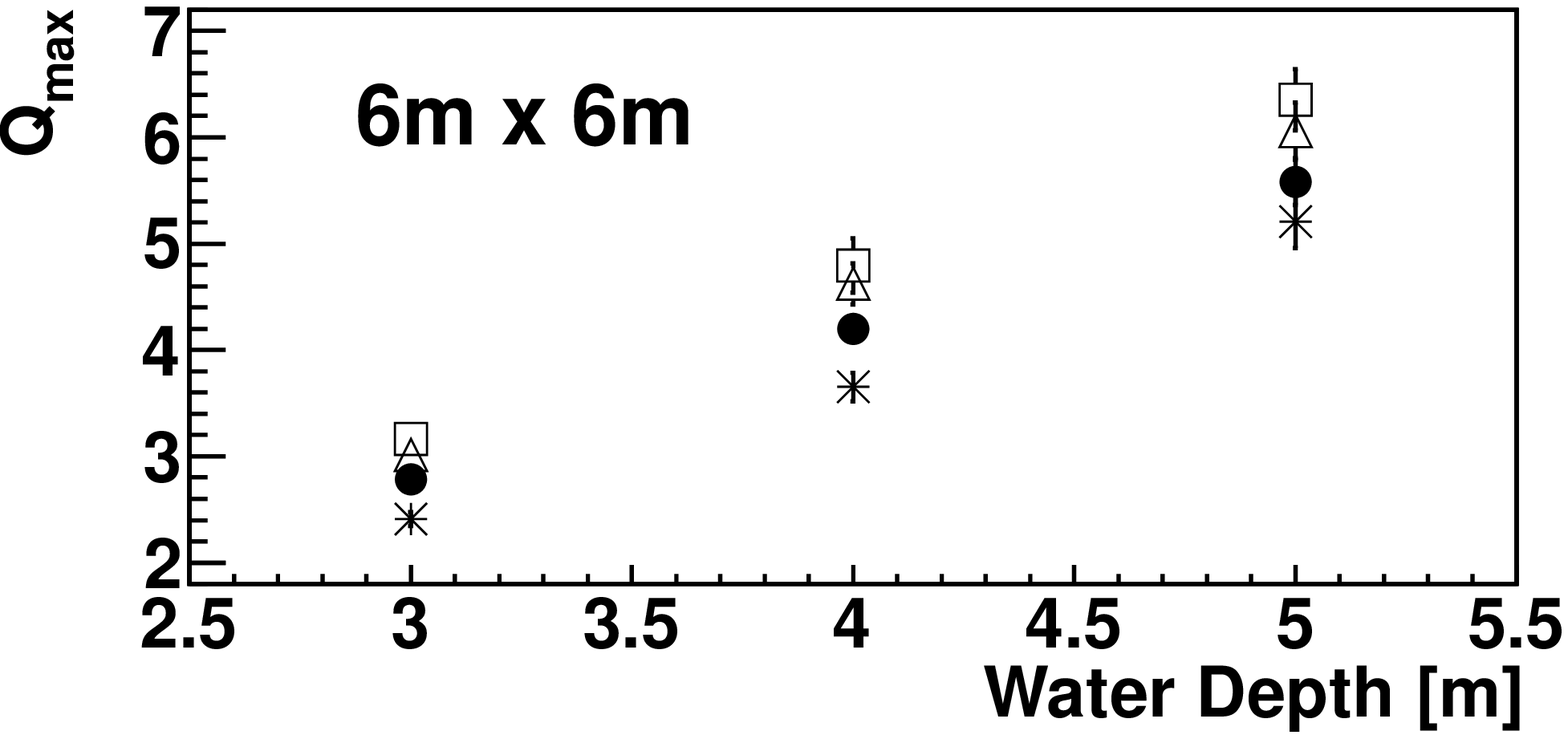}
\figcaption{\label{fig7} The $Q_{\rm max}$ for 36 configurations of the array.}
\end{center}

\subsubsection{Angular resolution}

Only gamma showers are meaningful for this analysis. With the hit information, using the official program developed for water Cherenkov arrays, the shower direction is reconstructed. Selecting events with the compactness threshold used by the $Q$ optimization, from the distribution of the opening angle between the reconstructed and the real directions, the angular resolution  is obtained. The Rayleigh distribution is assumed so that the Gaussian-like standard deviation~\cite{lab10} is assigned. Plots in figure 8 show values $1/\theta$ for different configurations. Similar trend as efficiency plots (figure 5) is observed, because the angular resolution is heavily dependent on the number of hits, which is actually determined by the efficiency.

\begin{center}
\includegraphics[width=0.70\linewidth]{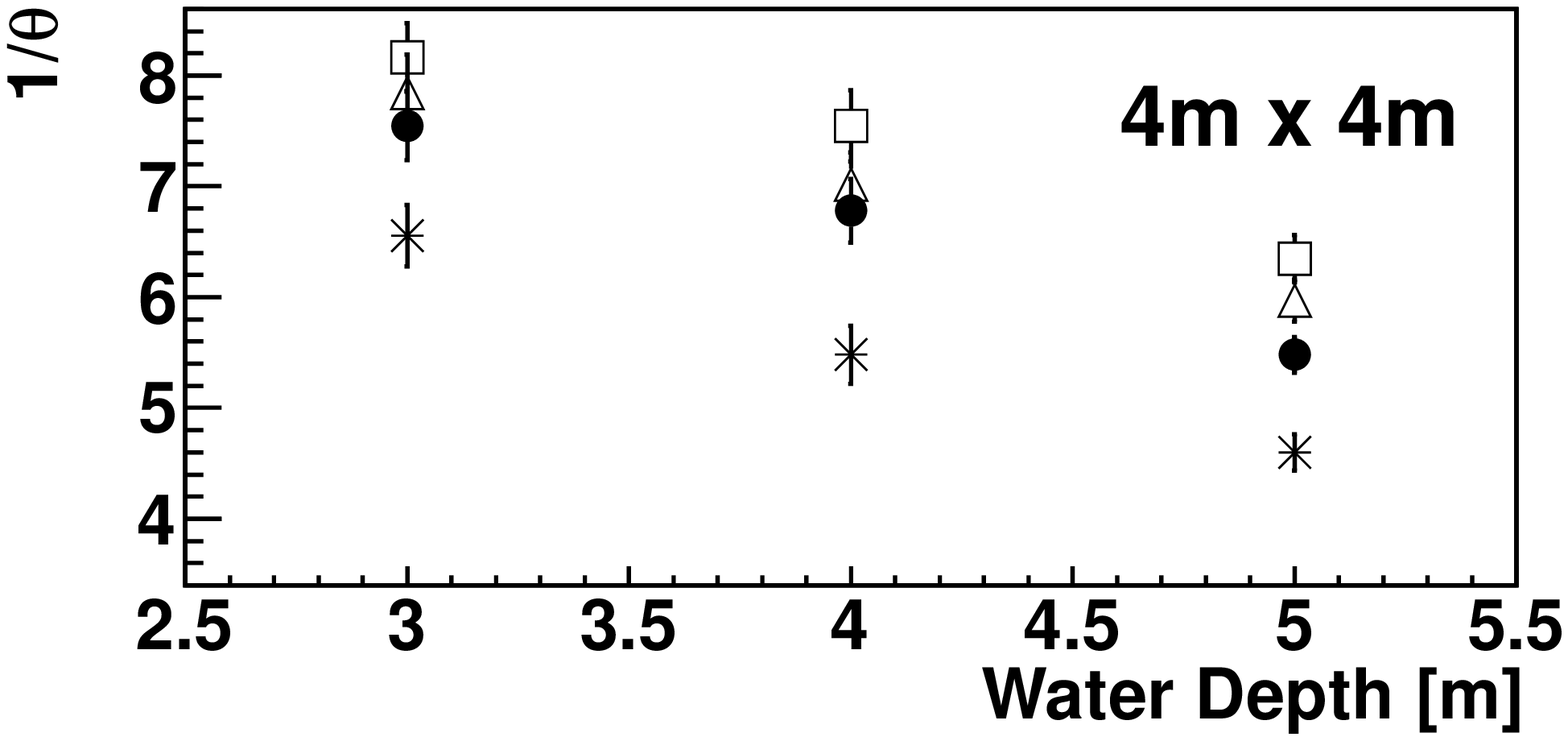}
\includegraphics[width=0.70\linewidth]{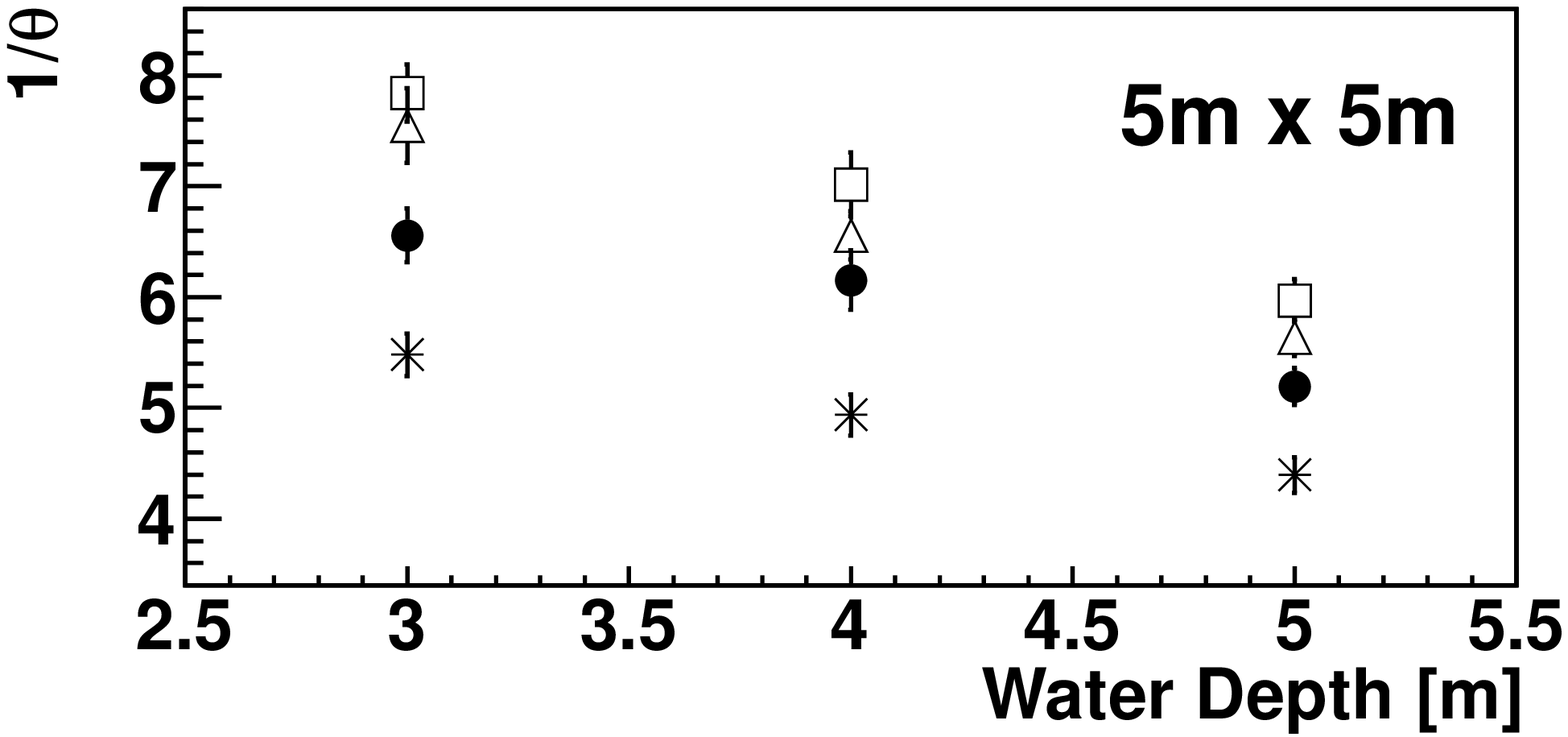}
\includegraphics[width=0.70\linewidth]{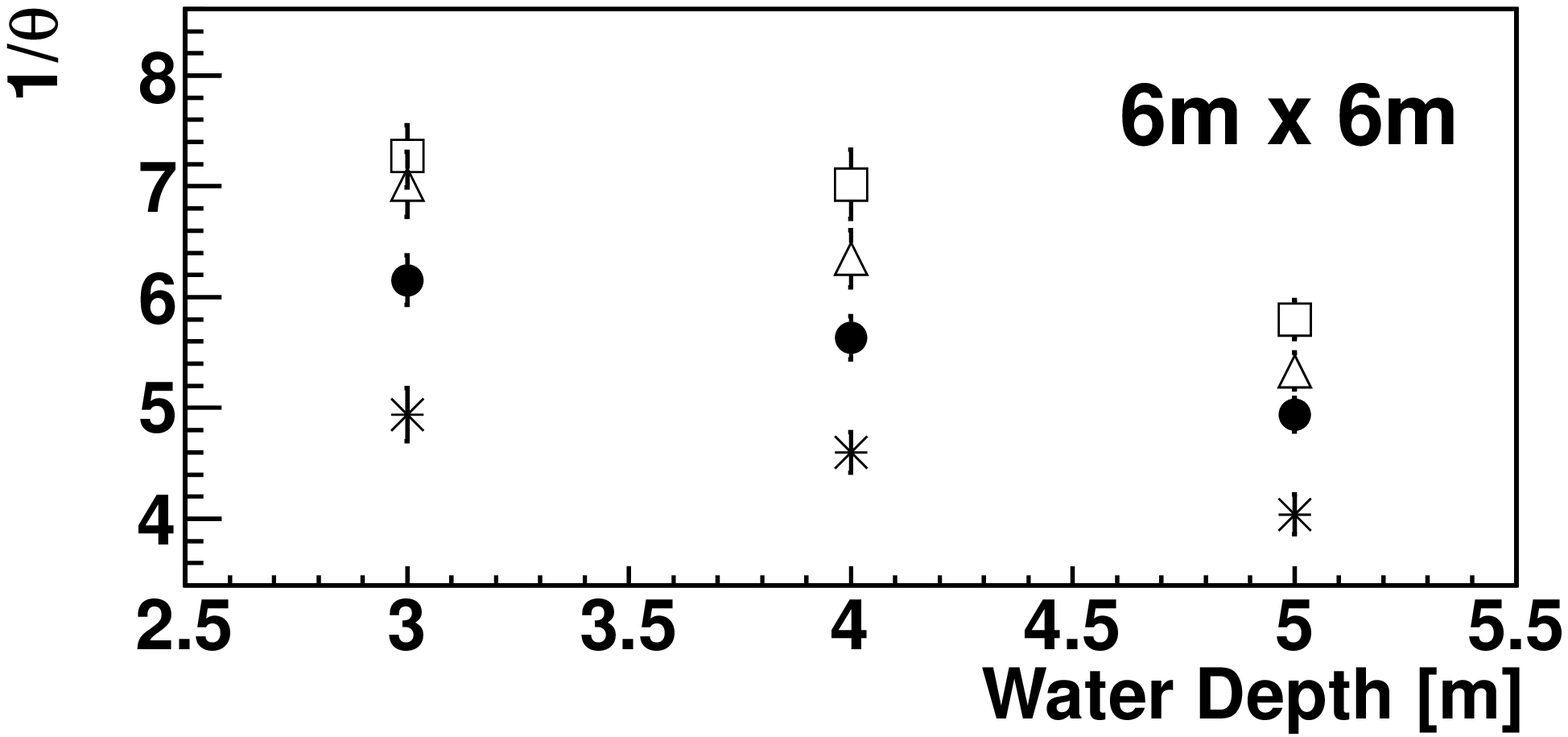}
\figcaption{\label{fig8} Inverse angular resolution $1/\theta$ for 36 configurations of the array.}
\end{center}

\subsubsection{Cost performance}

Due to very weak dependence of angular resolution on the compactness threshold around the optimized value, we have not bothered to optimize the performance with two factors in equation 3 in company. Tests to two cases of 36 configurations show also no difference at all.
The performance and cost performance is finally calculated for the 36 configurations with equation 1 and 3, whose results is shown in figure 9, 10. Controlled by the $Q_{\rm max}$, both the performance and cost performance turn better at deeper water depth, totally contrasting to the trend in approach $i$. If taking the statistical errors into account, configurations with less PMTs in a cell, e.g., 1 or 2, have better cost performance. The cell size affects the cost performance too, but the influence for cases with 1 or 2 PMTs in a cell is marginal.

\begin{center}
\includegraphics[width=0.70\linewidth]{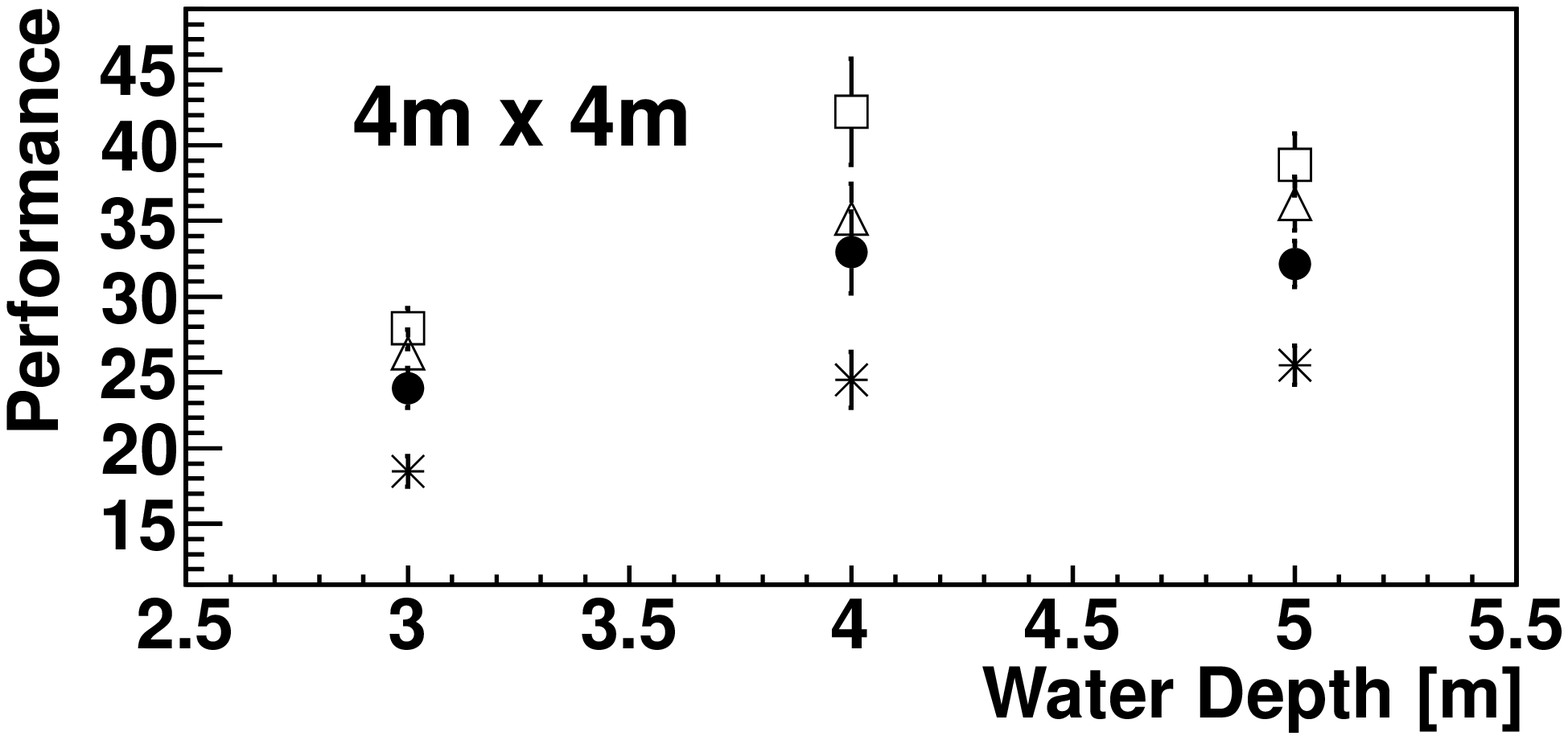}
\includegraphics[width=0.70\linewidth]{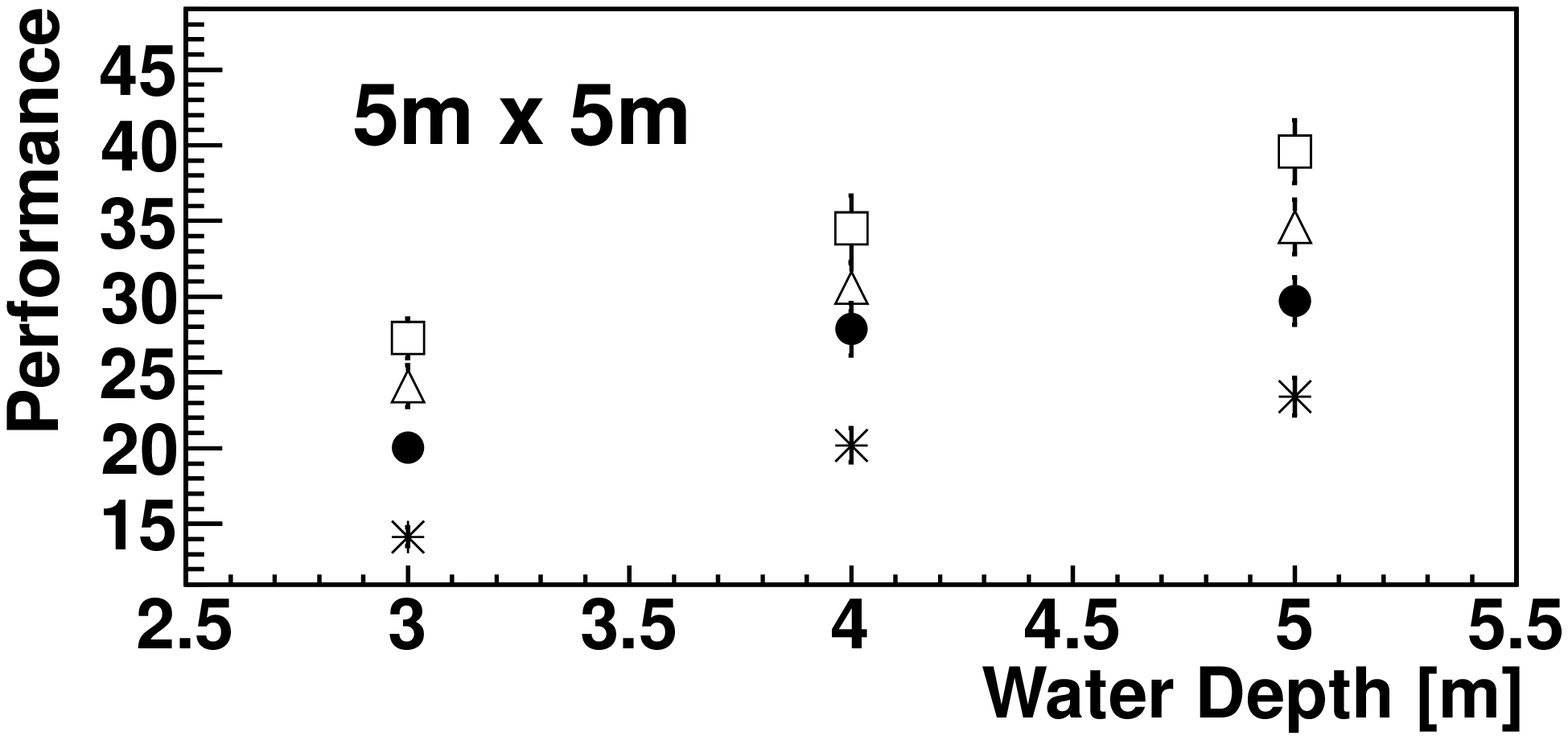}
\includegraphics[width=0.70\linewidth]{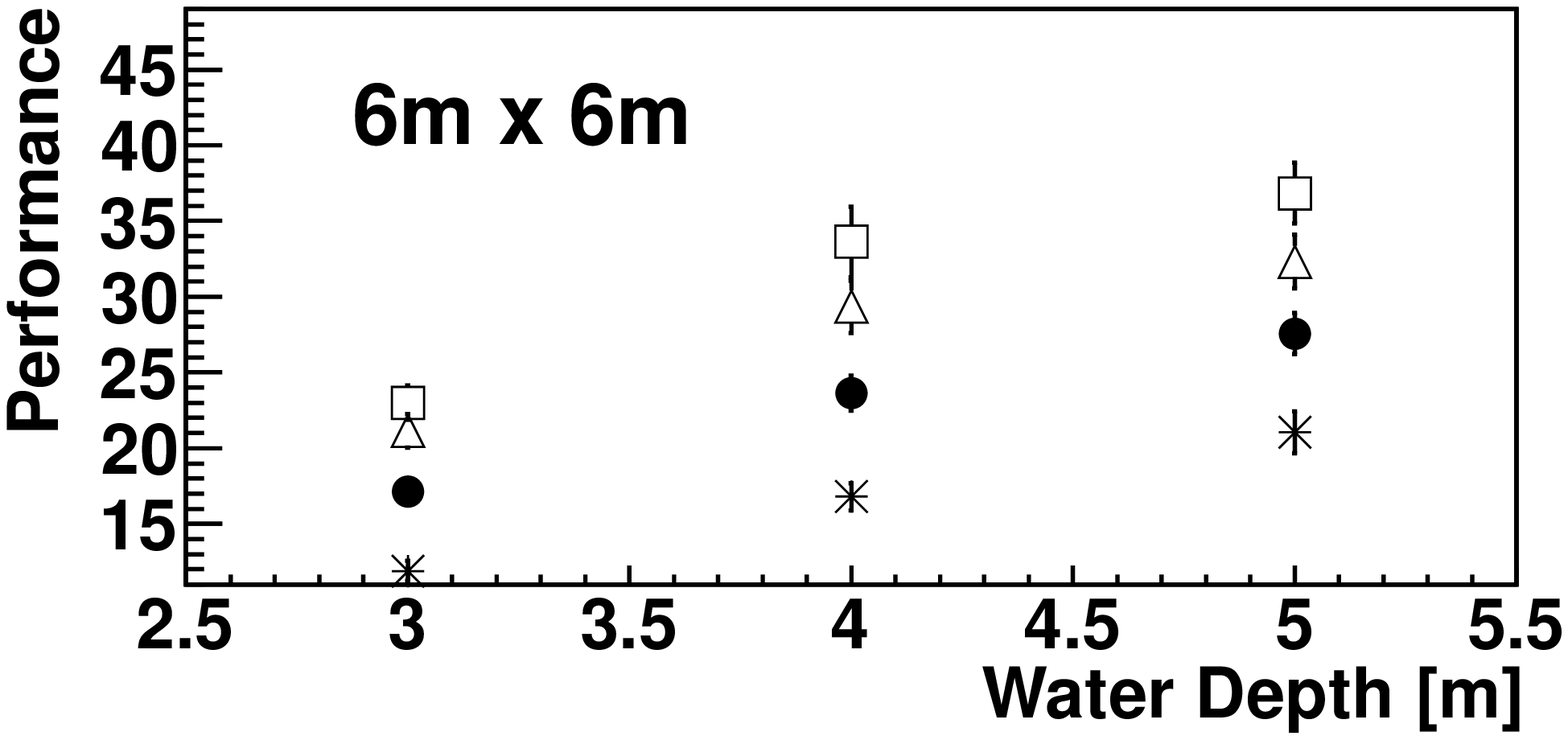}
\figcaption{\label{fig9}  Performance for 36 configurations of the array for the energy of 1 TeV gamma and 2 TeV proton.}
\end{center}

\begin{center}
\includegraphics[width=0.70\linewidth]{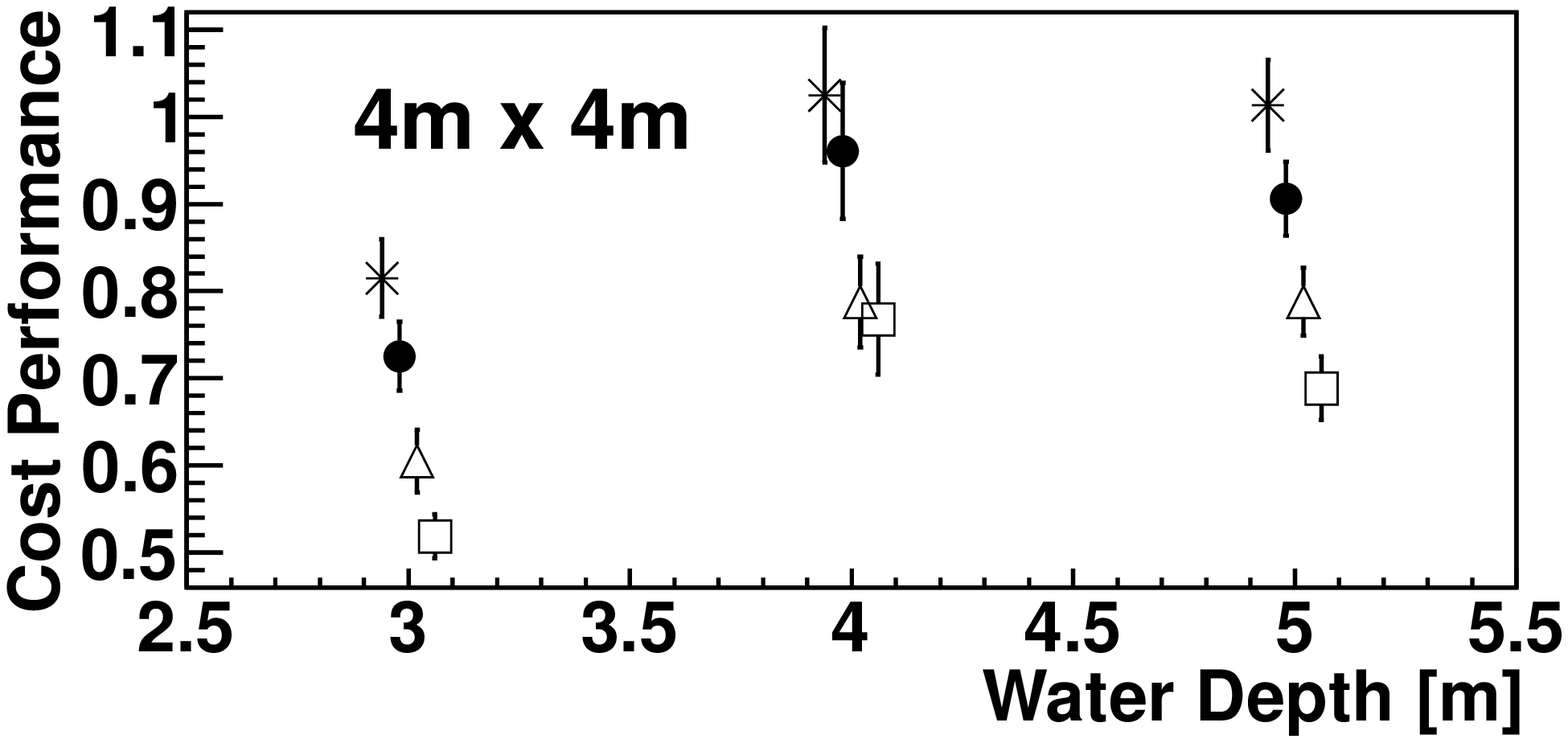}
\includegraphics[width=0.70\linewidth]{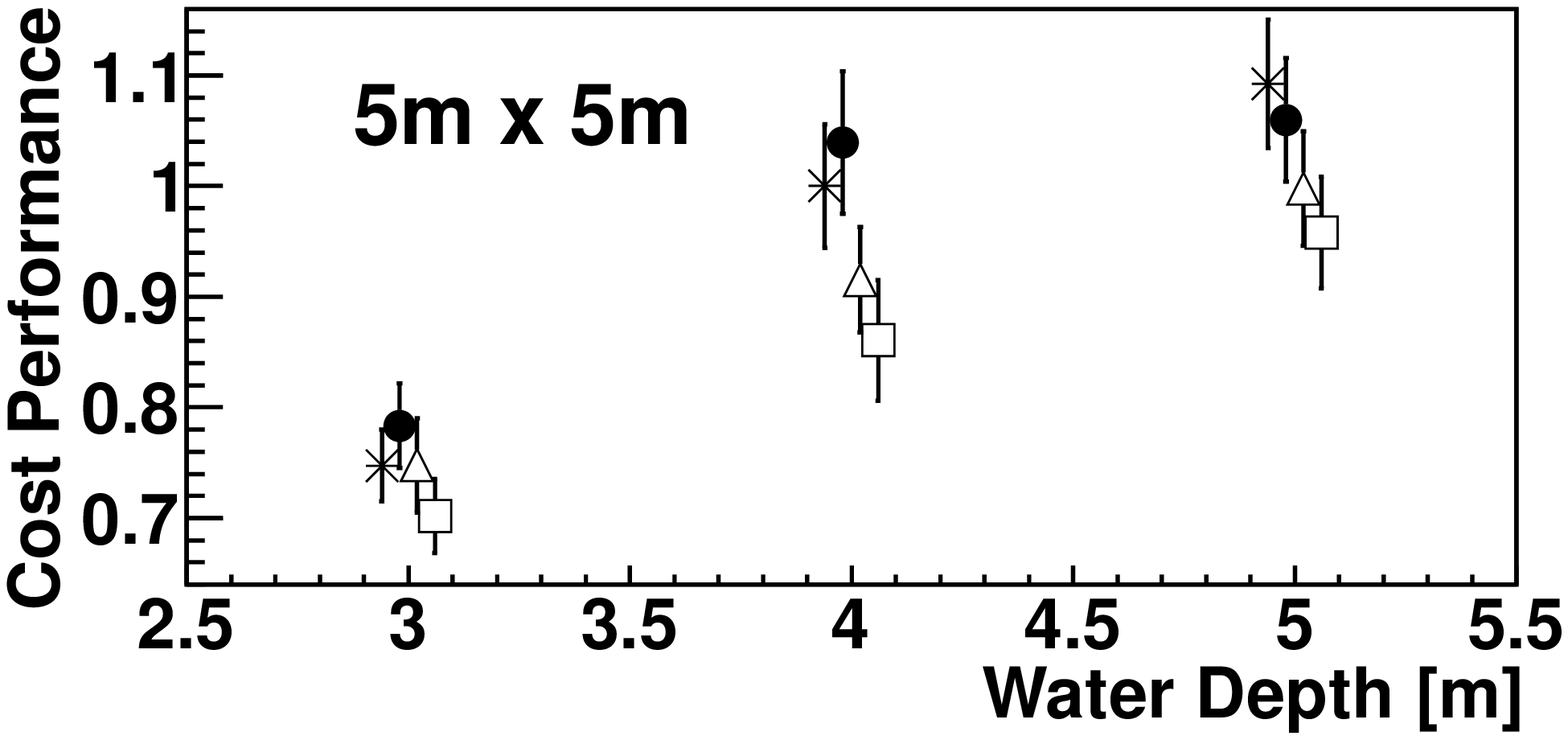}
\includegraphics[width=0.70\linewidth]{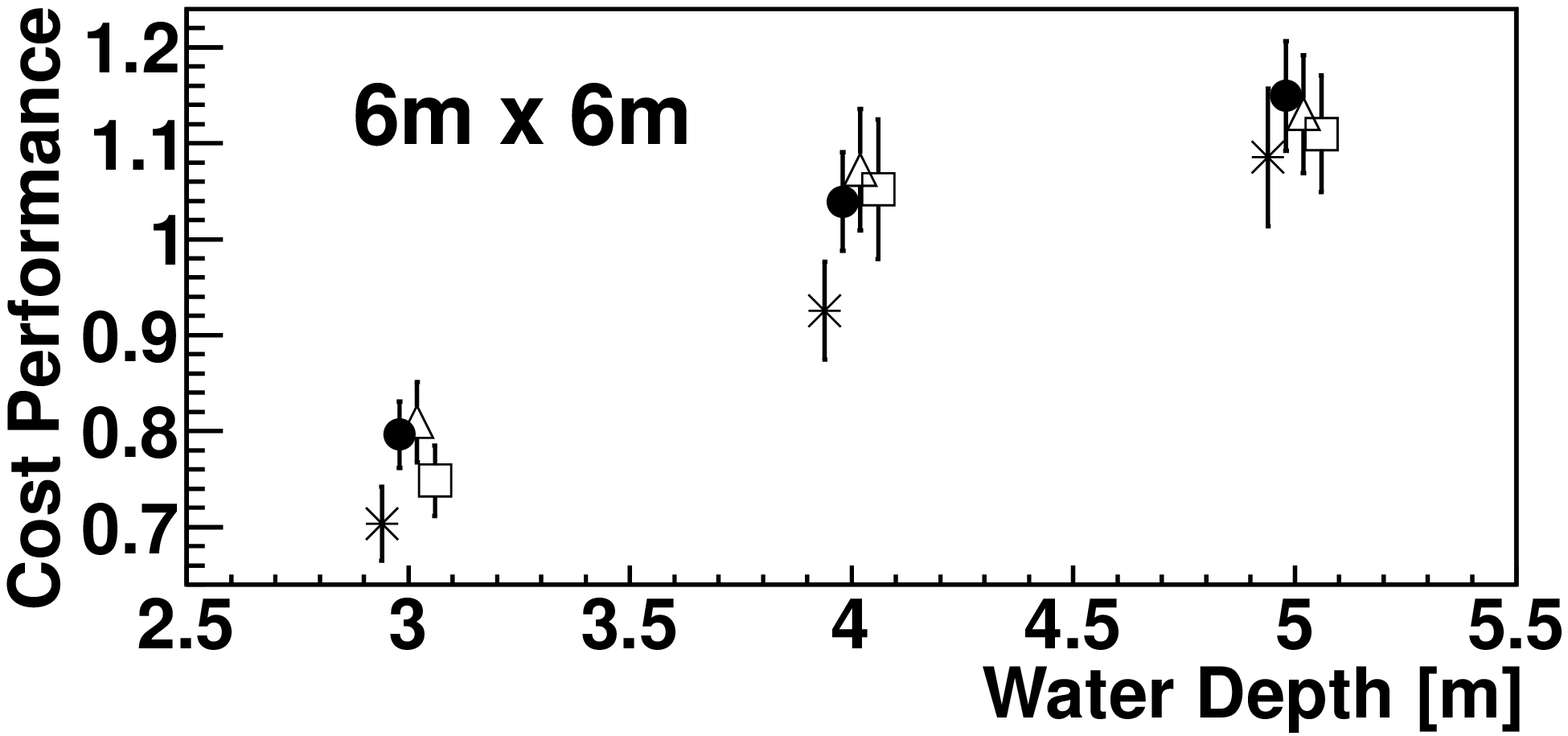}
\figcaption{\label{fig10} Cost performance for 36 configurations of the array for the energy of 1 TeV gamma and 2 TeV proton.}
\end{center}

\section{Discussions and conclusion}

Besides only simple scenarios are investigated, there are still some other realistic situations not being considered yet. For example, for the water quality, the absorption length 27~m at 400~nm might be too good to be maintained. If an absorption length 15~m is used, the middle plot in figure 10 would look like the one in figure 11 - the rising trend going with the water depth is depressed. Another issue is the accidentally coincident muon, which is not taken into account in the simulation. For configurations with several PMTs in a cell, the muon may fire all these PMTs altogether, troubling the reconstruction and even the trigger.
\begin{center}
\includegraphics[width=0.70\linewidth]{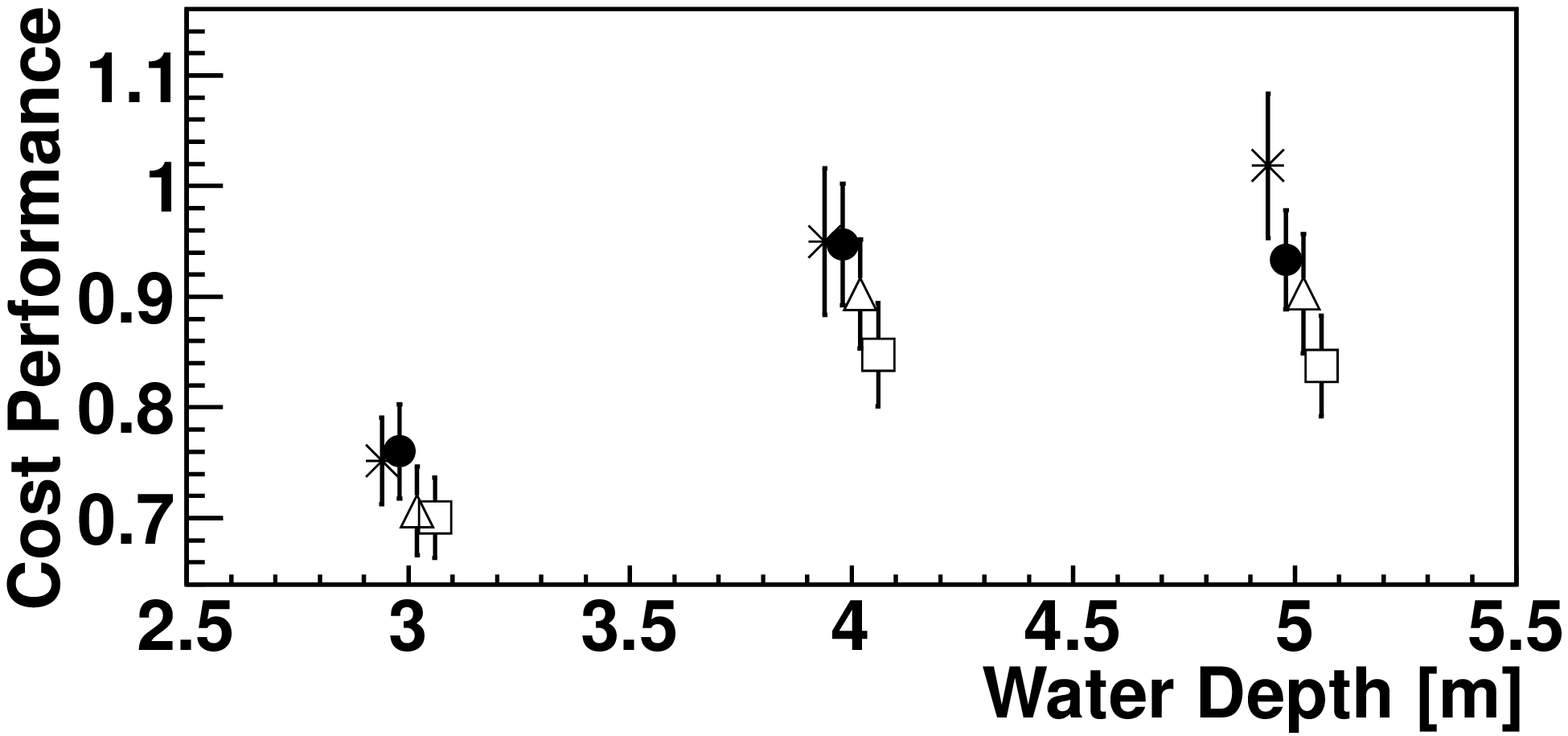}
\figcaption{\label{fig11} Cost performance for configurations with cell size 5~m $\times$ 5~m, in case that water absorption length is 15~m at 400~nm.}
\end{center}

Bigger cell size is more preferred by the simulation, no matter which approach, $i$ or $ii$. But the difference between 5~m $\times$ 5~m and 6~m $\times$ 6~m is trivial, with an effect $<10\%$. In practice, a cell size in between can be chosen out of engineering considerations.

Contrasting results on water depth optimization are found between the two approaches. As what has mentioned in section 2, these two approaches actually focus on gamma rays at different energy ranges. At low energies, gamma proton discrimination with compactness analysis does not work very well, so the angular resolution, i.e., efficiency, dominates in the performance. This point is proved by the simulation of lower primary energies in approach $ii$, where gamma is 0.5~TeV and proton is 1~TeV, and the result shows that the 4~m rather than 5 m water depth is the best. Considering that this kind of water Cherenkov detector array is not very sensitive at low energies such as below 1~TeV~\cite{lab4}, it is more proper to optimize the cost performance towards higher energies - in another word, deeper water depth is more preferred. But at the same time, the water quality issue should not be ignored. A water depth in between 4~m and 5~m shall be appropriate.

No critical difference on PMT quantity selection is found in two approaches.
One PMT in a cell shall be the best option.

In summary, based on this simulation work, a configuration of cell size in between 5~m $\times$ 5~m and 6~m $\times$ 6~m, water depth in between 4~m and 5~m, and 1 PMT in a cell is the best in cost performance for the water Cherenkov detector array of LHAASO.

\section*{Acknowledgments}

The authors would like to express their gratitude to the Milagro Collaboration for beneficial discussions on their experiences. This work is partly supported by NSFC (11175147) and the Knowledge Innovation Fund of IHEP, Beijing.\\

\end{multicols}

\clearpage


\begin{thebibliography}{90}

\vspace{3mm}

\bibitem{lab1} Weeks,~T.~C. {\it et al\/}., {\it ApJ} {\bf 342} (1989) 379--395.
\bibitem{lab2} Hinton,~J. Ground-based gamma-ray astronomy with Cherenkov telescopes, New J. Phys. 11 (2009) 055005.
\bibitem{lab3} Sinnis,~G.  Air shower detectors in gamma-ray astronomy, New J. Phys. 11(2009) 055007.
\bibitem{lab4} Yao,~Z.~G. {\it et al\/}., for the LHAASO Collaboration, {\it Proceedings of 32th ICRC} (2011).
\bibitem{lab5} Cao,~Z. A future project at tibet: the large high altitude air shower observatory (LHAASO), CHINESE PHYSICS C,2010,34(2):4,249-252.
\bibitem{lab6} {\tt http://www-ik.fzk.de/{\textasciitilde}corsika}.
\bibitem{lab7} Ostapchenko,~S.~S. Nucl.phys.B (Proc. Suppl.)151(2006)143-147;Phys.Rev.D 74 (2006) 014026.
\bibitem{lab8} Agostinelli,~S. {\it et al\/}., {\it Nucl. Instru. \& Meth. in Phys. Res.} {\bf A 506} (2003) 250--303. See also {\tt http://geant.cern.ch/}.
\bibitem{lab9} {\tt http://neutrino.phys.ksu.edu/{\textasciitilde}GLG4sim}.
\bibitem{lab10} The Review of Particle Physics, J. Beringer et al. (Particle Data Group), Phys. Rev. D86, 010001 (2012).
\end{thebibliography}
\end{document}